\documentclass[manuscript]{aastex}

\shorttitle{Light Brigade in Developing AR. I.}
\shortauthors{Toriumi et al.}

\begin{document}


\title{Light Bridge in a Developing Active Region.\\
  I. Observation of Light Bridge
  and its Dynamic Activity Phenomena}


\author{Shin Toriumi$^{1}$, Yukio Katsukawa$^{1}$, and Mark C. M. Cheung$^{2}$}
\affil{$^{1}$National Astronomical Observatory of Japan, 2-21-1 Osawa, Mitaka, Tokyo 181-8588, Japan}
\email{shin.toriumi@nao.ac.jp}
\affil{$^{2}$Lockheed Martin Solar and Astrophysics Laboratory, 3251 Hanover Street, Building/252, Palo Alto, CA 94304, USA}




\begin{abstract}
Light bridges,
the bright structures
that divide the umbra of sunspots and pores
into smaller pieces,
are known to produce
wide variety of activity events
in solar active regions (ARs).
It is also known that
the light bridges appear
in the assembling process
of nascent sunspots.
The ultimate goal
of this series of papers
is to reveal the nature
of light bridges
in developing ARs
and the occurrence
of activity events
associated with the light bridge structures
from both observational and numerical approaches.
In this first paper,
exploiting the observational data obtained
by {\it Hinode}, IRIS,
and {\it Solar Dynamics Observatory} ({\it SDO}),
we investigate
the detailed structure
of the light bridge
in NOAA AR 11974
and its dynamic activity phenomena.
As a result,
we find that
the light bridge has
a weak, horizontal magnetic field,
which is transported
from the interior
by large-scale convective upflow
and is surrounded
by strong, vertical fields
of adjacent pores.
In the chromosphere
above the bridge,
a transient brightening occurs
repeatedly and intermittently,
followed by
a recurrent dark surge ejection
into higher altitudes.
Our analysis indicates that
the brightening is the plasma heating
due to magnetic reconnection
at lower altitudes,
while the dark surge
is the cool, dense plasma
ejected from the reconnection region.
From the observational results,
we conclude that
the dynamic activity
observed in a light bridge structure
such as chromospheric brightenings
and dark surge ejections
are driven by magnetoconvective evolution
within the light bridge
and its interaction
with surrounding magnetic fields.
\end{abstract}





\section{Introduction
  \label{sec:introduction}}

Solar active regions (ARs)
are known to produce
a wide variety of dynamic phenomena
from sub-arcsec events like
Ellerman bombs \citep{ell17}
to catastrophic eruptions like
solar flares and coronal mass ejections
\citep{shi11}.
It is now widely accepted that
such phenomena
are the result
of magnetic reconnection,
the dynamic physical mechanism
that changes magnetic topology
and releases magnetic energy.

It is also known that
light bridges,
the bright structures
dividing the dark umbrae
of sunspots and pores
into multiple smaller elements,
produce many activity events
in ARs.
Light bridges typically have
a weaker magnetic field
compared to surrounding umbrae
with field lines highly inclined
and aligned along the long axis of the bridges
\citep{bec69,lit91,rue95,lek97}.
The magnetic fields around the light bridge
have a canopy structure \citep{jur06},
which
supports the
theoretical model
that the light bridges are
the ``field-free''
intrusion of hot plasma
into the umbral magnetic field
\citep{par79b,cho86,spr06}
and recent numerical simulations
of sunspots
that apply radiative magneto-convection
\citep{rem09,che10,rem11}.
The canopy configuration
may be responsible for
dynamic phenomena
observed in the higher atmosphere
such as \ion{Ca}{2} H brightenings and ejections
\citep{lou08,shi09,lou09},
H$\alpha$ surges
\citep{roy73,asa01,bha07},
brightenings in the
{\it TRACE} 1600 {\AA} channel
\citep{ber03},
etc.
Regarding the light bridge formation,
\citet{kat07} observed
a mature sunspot
with high spatial resolution
and found that
many umbral dots appear
in the umbra of the spot.
They rapidly intrude
into the umbra
and eventually form
a light bridge
that cuts across
the umbra.

To a large extent,
studies of light bridges
focused on the fragmentation phase
of decaying sunspots.
However,
it is known that
light bridges appear
also in the assembling process
of nascent sunspots
\citep{bra64,gar87}.
In this case,
bridges are created
between merging pores
that eventually grow
into a single spot.
It is highly possible that
the formation
of such light bridges
is closely connected
to the large-scale flux emergence
from the convection zone.
From this point of view,
the light bridge structure
in a developing AR
is not only a driver
of various dynamic activity phenomena
but also one of the most important keys
to understand
the flux emergence
and the resultant spot formation.

The aim of this series of papers
(this paper along with \citet{tor15b},
hereafter Paper II\defcitealias{tor15b}{Paper~II})
is to reveal
the nature of light bridges
in newly emerging ARs
and the occurrence
of various activity events
related to the light bridge structures
from both observational
and numerical perspectives.
In the present paper,
the first paper of the series,
we analyze
observations of NOAA AR 11974
to study the detailed structure
of the light bridge
that appeared
during the course
of spot formation
and its relation with
dynamic events
that occurred
in and around the light bridge.
The organization of this paper
is as follows.
In Section \ref{sec:observations},
we describe the data that we used
and give an overview of this AR.
Then,
in Sections \ref{sec:results1}
and \ref{sec:results2},
we show the analysis results
of the light bridge
and activity phenomena,
respectively,
which is followed by
the summary of observational results
in Section \ref{sec:summary}.
We discuss the results
in Section \ref{sec:discussion}
and, finally,
we conclude the paper
in Section \ref{sec:conclusion}.
Based on the observational results
of the present work,
in \citetalias{tor15b},
we search for the formation process
of a light bridge
and the cause of activity events
by analyzing numerical simulation data
of AR formation
\citep{che10}.

\section{Observations
  \label{sec:observations}}

\subsection{Observations and Data Reduction}

For this analysis,
we utilized observation data
of the newly emerging
NOAA AR 11974,
which appeared on the southern hemisphere
in February, 2014.
In this subsection,
we describe the observation data
obtained by various instruments
and their reduction processes.

The large-scale evolution
in the photosphere is mainly covered
by Helioseismic and Magnetic Imager
(HMI; \citealt{sche12,scho12})
on board the {\it Solar Dynamics Observatory}
({\it SDO}; \citealt{pes12}).
We used the tracked cutout data
(line-of-sight (LOS) magnetogram,
Dopplergram, and intensitygram)
of AR 11974;
the field-of-view (FOV) ranges
$500\arcsec\times 250\arcsec$,
which is enough
for covering the entire AR.
The duration is
3 hours from 22:00 UT, February 13.
The pixel size and the time cadence
are $0\farcs 50$ and 45 s,
respectively.
From the Dopplergram
we subtracted
the effect of the solar rotation,
the orbital motion of the {\it SDO} satellite,
and the east-west trend due to the spherical geometry of the Sun
by applying the method
introduced in \citet{tor14}.
We also used the {\it SDO}/Atmospheric Imaging Assembly
(AIA; \citealt{lem12}) data
with the same FOV and duration
as those of the HMI data
to analyze the activity
in the upper atmosphere.
The data have
a pixel size of $0\farcs 60$
for all bands
and a cadence of 12 s 
for the extreme ultraviolet bands
(EUV: 94 {\AA}, 131 {\AA}, 171 {\AA}, 193 {\AA},
211 {\AA}, 304 {\AA}, and 335 {\AA}),
24 s for the two ultraviolet bands
(UV: 1600 {\AA} and 1700 {\AA}),
and 3600 s for the continuum band (4500 {\AA}).
Both HMI and AIA data
were exported
from the Joint Science Operations Center
(JSOC)\footnote{http://jsoc.stanford.edu/}.
They were mapped onto a common grid
using the {\tt aia\_prep} routine
in the {\it SolarSoftWare} package
\citep{fre98}.

During the period of the above-mentioned {\it SDO} data,
the Solar Optical Telescope (SOT; \citealt{tsu08})
on board the {\it Hinode} satellite \citep{kos07}
tracked this AR
and obtained \ion{Ca}{2} H (3968.5 {\AA})
broadband filtergram images
with the FOV of $223\farcs2\times 111\farcs 6$.
The time cadence and the pixel size
of this data
are 1 minute and $0\farcs218$,
respectively.
We calibrated the Ca data
with the {\tt fg\_prep} procedure
for dark-current subtraction and flat fielding.
We also used the
raster data
of the spectropolarimeter
(SP; \citealt{lit13}:
two magnetically sensitive \ion{Fe}{1} lines
at 6301.5 {\AA} and 6302.5 {\AA}).
In this time slot,
there was one raster scan
that starts at 00:00 UT
and ends at 00:55 UT, February 14.
The FOV is $255\farcs8\times 122\farcs9$
with the spatial sampling of $0\farcs30$
and the step size of $0\farcs32$.
In this study
we used the SOT/SP level 2 data,
which are the outputs
from the Milne–Eddington gRid Linear Inversion Network code
(MERLIN; \citealt{lit07}).
Note that
the correction
of orbital variation
is done
when the level 2 data
are generated.
The 180$^{\circ}$ ambiguity
in the azimuth angle
of the magnetic field
is resolved
by using the AZAM utility
\citep{lit95}.

Along with {\it Hinode},
the Interface Region Imaging Spectrograph
(IRIS; \citealt{dep14})
was also monitoring AR 11974
during this period.
Here we used the IRIS data
made from 22:14 UT, February 13,
to 00:47 UT, February 14.
This flarewatch program
(OBS 3860259280)
consists of repeated 8-step rasters
with a step size of $2\farcs0$.
During the observation period
there were 200 scans
for the near-ultraviolet
(NUV: 2790--2834 {\AA})
and far-ultraviolet
(FUV: 1333--1356 {\AA} and 1399--1406 {\AA})
wavelength bands
and the first 122 scans were available.
The pixel size is $0\farcs 167$
and the FOV for each scan
is $14\farcs 1\times 119\farcs 4$.
The step cadence and the raster cadence
are 9.4 s and 75 s,
respectively.
At the same time,
slit-jaw images (SJIs)
in the filters
of 1330 {\AA} and 1400 {\AA}
were taken
at a cadence of 19 s.
The FOV is
about $118\arcsec\times 118\arcsec$.
We used the level 2 data,
which take into account
the dark-current subtraction,
flat fielding,
and geometrical correction.

The co-alignment among various observation data sets
were achieved
by using AIA images
as a reference:
For the HMI and SOT/SP data
cross-correlations
between the AIA 4500 {\AA} image
and the HMI intensitygram and SP Stokes-I image
were used,
while for the IRIS and SOT Ca data
cross-correlations
between the AIA 1600 {\AA} images
and the IRIS 1330 {\AA}, 1400 {\AA}, and Ca images
were used.

\subsection{Evolution of AR 11974
  and Overview of Activity Phenomena
  \label{subsec:longterm}}

\begin{figure}
  \begin{center}
    \includegraphics[width=160mm]{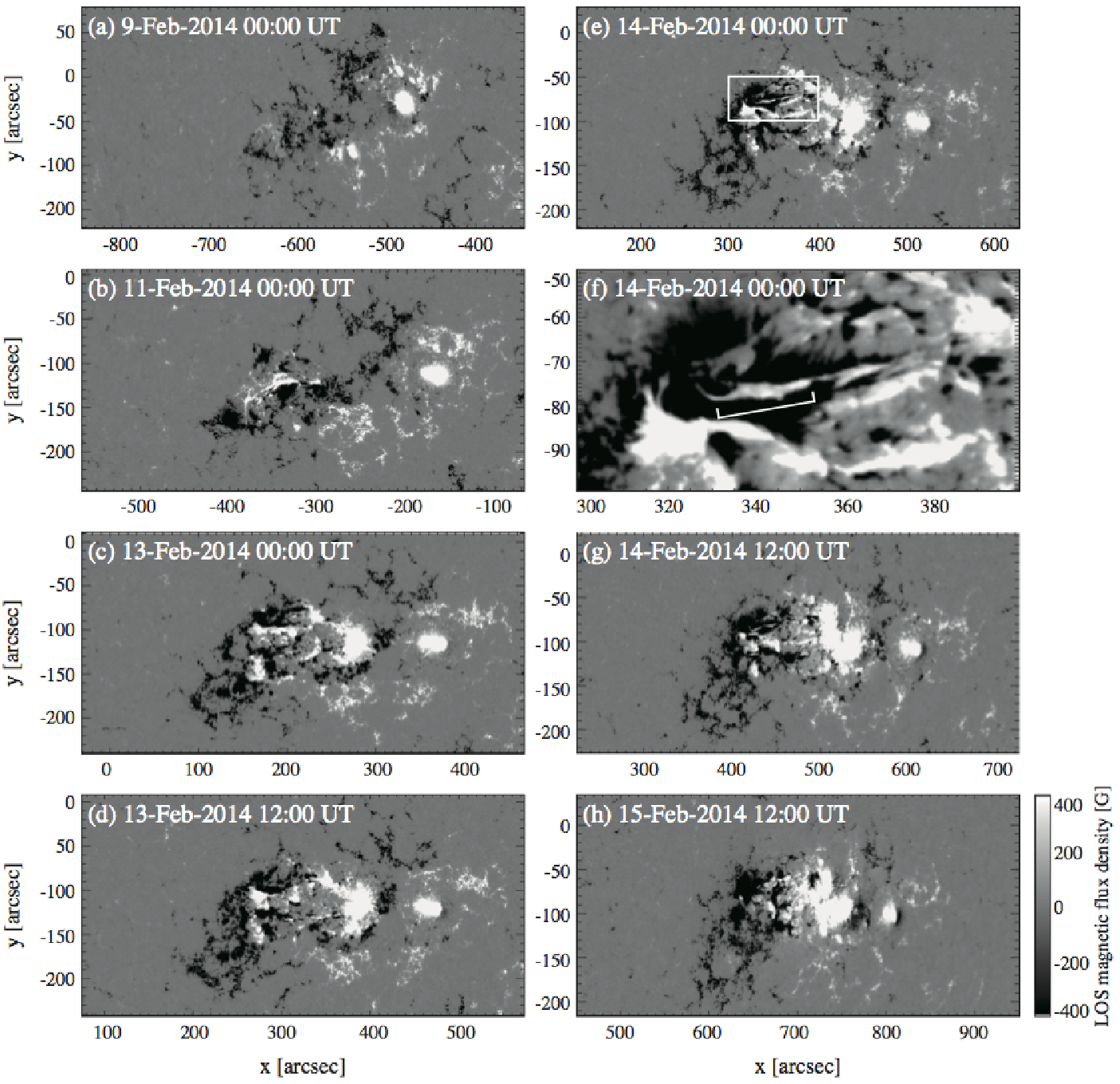}
  \end{center}
  \caption{LOS magnetogram
    of NOAA AR 11974
    obtained by {\it SDO}/HMI.
    White box in panel (e)
    indicates the FOV of panel (f).
    The light bridge structure
    sandwiched between two negative pores
    is shown in panel (f).
    In this figure,
    $(x, y)$ represent
    the heliocentric-Cartesian coordinates,
    where the origin $(0, 0)$ is located
    at the disk center.}
  \label{fig:longterm}
\end{figure}

Figure \ref{fig:longterm} shows
the long-term evolution
of NOAA AR 11974.
This AR appeared
on the southeastern limb
on 2014 February 5
and, since then,
it showed
significant flux emergence
to the east of the pre-existing positive sunspot
and formed a complex AR
(panels (a) to (d)).
From February 13,
additional flux emerged
within the AR,
followed by the formation
of a light bridge
in the following polarity proto-spot
(panel (e)).
Panel (f)
is the enlarged version of (e),
which shows that
an elongated structure
(i.e. light bridge)
has a weak positive LOS magnetic field
and is sandwiched between
two negative pores.
The bridge has a length
(extension along the long axis)
of $\sim 30\arcsec\sim 22\ {\rm Mm}$
with a width
(thickness across the bridge)
of $\sim 4\arcsec\sim 3\ {\rm Mm}$.
Throughout the evolution,
emerged small-scale magnetic elements
of the same polarity
gradually coalesced into
larger magnetic elements.
In a similar manner,
the two negative pores
around the light bridge
came closer to merge with each other
and thus the bridge became squeezed out
(panel (g)).
On February 15,
as shown in panel (h),
the magnetic elements of this AR
eventually built up
two major sunspots
of positive and negative polarities
and, until this moment,
the light bridge structure
had completely disappeared.
Thus,
the lifetime of this light bridge
is about 2 days.

Together with flux emergence,
the target AR 11974 showed
a wide variety of
dynamic phenomena.
Figure \ref{fig:tile} overviews
the activity events
in this AR.
In panels (i) and (j),
the light bridge structure
is clearly seen
between the negative polarity pores.
This light bridge basically shows
a blueshift indicating upward motion
(panel (k)).
In channels that image
the photosphere and chromosphere
(e.g., AIA 1600 {\AA} and 1700 {\AA},
IRIS 1330 {\AA} and 1400 {\AA},
and SOT \ion{Ca}{2} H
as in panels (d--h)),
small-scale brightness enhancements
are found to
repeatedly occur
in the middle of the FOV
(see online movie).
Here we mark three representative brightenings
with A, B, and C.
In addition,
in the EUV images (panels (a--c)),
a dark surge (jet) recurrently extends upward
from the location of the light bridge,
particularly from brightening A
in the bridge
(see movie).
In the following sections,
we describe these features
in detail.
The light bridge structure
is analyzed in Section \ref{sec:results1},
while in Section \ref{sec:results2}
we investigate
the brightenings
and dark surges
and reveal the correspondence
with the bridge.

\section{Magnetic and Velocity Structures
  of the Light Bridge
  \label{sec:results1}}

The recurrent brightening events (A)
originate from positions
coincident with the light bridge.
Also,
the dark surges recurrently
stretch from the brightening events.
This motivates us
to investigate possible associations
between the light bridge evolution
and the brightening events.
We first examine
the detailed magnetic and velocity structures
of the light bridge.

\begin{figure}
  \begin{center}
    \includegraphics[width=140mm]{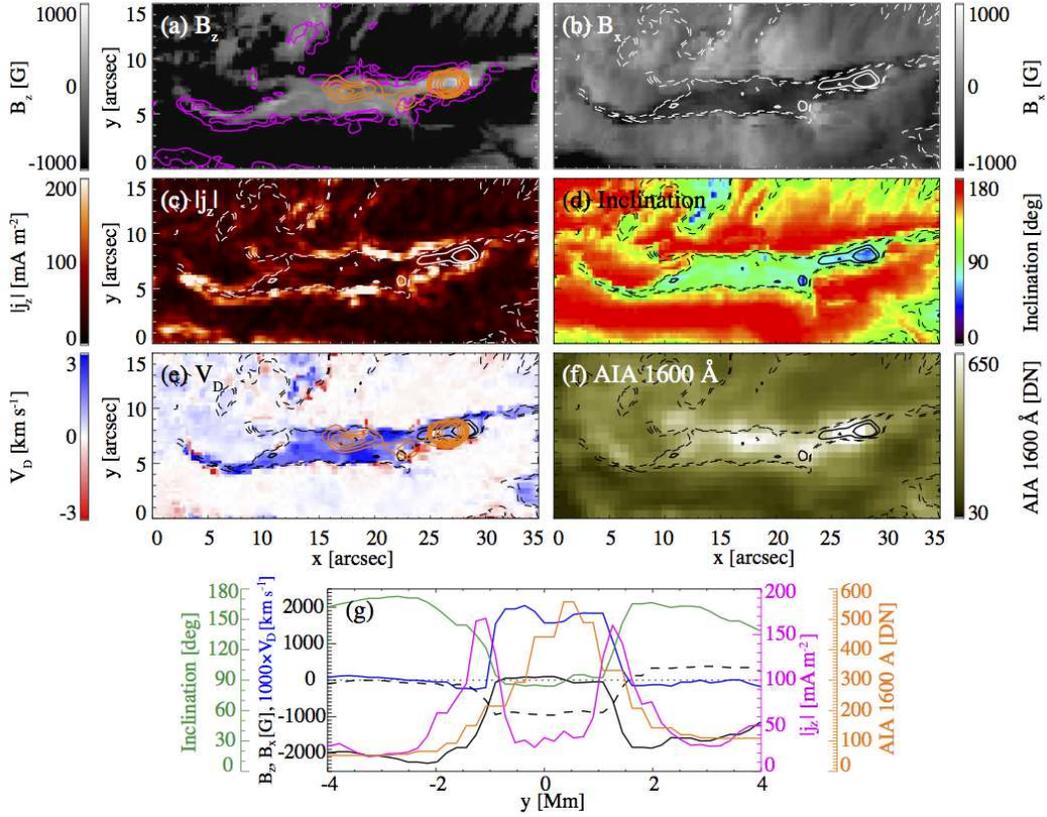}
  \end{center}
  \caption{
    {\footnotesize
    (a--e) SOT/SP data
    around the light bridge structure.
    This FOV is scanned westward
    from 00:16 to 00:23 UT,
    2014 February 14.
    (a) Vertical magnetic field strength $B_{z}$.
    The vertical electric current density $|j_{z}|$
    computed from $(B_{x}, B_{y})$
    is overlaid with purple contours.
    The contour levels are
    $|j_{z}|=100\ {\rm mA\ m}^{-2}$
    and $200\ {\rm mA\ m}^{-2}$.
    In addition,
    AIA 1600 {\AA} intensity
    at 00:20 UT
    is overlaid
    with orange contours.
    For the magnetic field,
    we use the local coordinates
    $(x,y,z)$,
    where $\hat{\mbox{\boldmath $z$}}$
    is in the local radial direction.
    (b) Horizontal field strength $B_{x}$.
    Solid and dashed contours show
    $B_{z}=400\ {\rm G}$ and $200\ {\rm G}$ levels
    and $B_{z}=-200\ {\rm G}$ and $-400\ {\rm G}$ levels,
    respectively
    (same for panels (c--f)).
    (c) Vertical current density $|j_{z}|$.
    (d) Inclination angle of the magnetic fields
    with respect to the local vertical.
    (e) Doppler (LOS) velocity $V_{\rm D}$,
    where positive (blue) and negative (red) signs
    indicate the upward and downward velocity,
    respectively.
    Orange contours indicate
    the AIA 1600 {\AA} intensity levels.
    (f) AIA 1600 {\AA} intensity
    at 00:20 UT.
    (g) One-dimensional ($y$-)profiles
    across the light bridge
    averaged over $15\farcs0\le x\le 22\farcs5$.
    Black solid, black dashed, blue, green, and purple lines indicate
    vertical field strength $B_{z}$,
    horizontal field strength $B_{x}$,
    Doppler velocity $V_{\rm D}$,
    inclination angle of the magnetic field,
    and vertical electric current density $|j_{z}|$,
    all obtained from SOT/SP.
    Orange line indicates
    the AIA 1600 {\AA} intensity
    at 00:20 UT, 2014 February 14.
    The origin of the $y$-axis
    is set at the middle
    between the two current ($|j_{z}|$) maxima,
    while the dotted line shows
    the $B_{z}=B_{x}=V_{\rm D}=0$ level.
    }
  }
  \label{fig:sp}
\end{figure}

Figures \ref{fig:sp}(a--e)
show the SOT/SP data
around the light bridge.
Panel (a) is the vertical magnetic field strength
$B_{z}$.
For mapping the magnetic field,
we use the local Cartesian reference frame $(x,y,z)$
with $\hat{\mbox{\boldmath $z$}}$ being
the local radial direction.
In Panel (a),
the vertical field
is overlaid by
the absolute vertical electric current density
contours
$|j_{z}|$
(corresponding to panel (c)),
where
\begin{eqnarray}
  |j_{z}|=
  \left|
    \frac{1}{\mu_{0}}
    \left(
      \frac{\partial B_{y}}{\partial x}
      - \frac{\partial B_{x}}{\partial y}
    \right)
  \right|
\end{eqnarray}
and $\mu_{0}$ is
the magnetic permeability,
and the AIA 1600 {\AA} intensity contours
(corresponding to panel (f)).
Panels (b--d) are
the maps for
the horizontal field $B_{x}$,
the vertical current $|j_{z}|$,
the inclination angle
with respect to the local vertical.
Panel (e) shows
the Doppler (i.e. LOS) velocity $V_{\rm D}$.
In order to calibrate
the absolute velocity,
we subtracted the mean Doppler velocity
calculated from the quiet Sun
in the same SP data.
Panel (f) shows
the AIA 1600 {\AA} intensity.

From Figures \ref{fig:sp}(a--d),
we can see that
the light bridge has
an almost horizontal magnetic field
with a lower strength
($B_{z}\sim 0\ {\rm G}$
and $B_{x}\sim -1000\ {\rm G}$)
at the photosphere
and is sandwiched
between the strong,
mostly vertical negative pores
($B_{z}\sim -2000\ {\rm G}$).
We can also see that
the enhancements
of the vertical current
($|j_{z}|\gtrsim 100\ {\rm mA\ m}^{-2}$)
are concentrated
at the edges of the bridge.
Photospheric magnetograms
in a single layer
do not allow us
to compute horizontal components
of the current
to obtain the absolute current density
$|\mbox{\boldmath $j$}|
=|\nabla\times\mbox{\boldmath $B$}|/\mu_{0}$.
Nevertheless,
the observed distribution
of $|j_{z}|$ suggests
the presence of magnetic shear,
which may be favorable
for magnetic reconnection.
It should be noted here that,
since the magnetic fields are
observed on the surface
of a constant optical depth
(iso-$\tau$ surface),
$|j_{z}|$ may be affected
by the vertical fluctuation
of this surface.
In \citetalias{tor15b},
we analyze the simulation data
and find that this effect
is relatively small.

Figure \ref{fig:sp}(e) tells us
that almost the entire light bridge
shows upward motion
($V_{\rm D}\gtrsim 1\ {\rm km\ s}^{-1}$).
In contrast,
the edge of the bridge shows
a downflow,
which is much narrower
and faster
($V_{\rm D}$ down to $-6\ {\rm km\ s}^{-1}$).
Inside of the bridge,
we can see
localized patches
of enhanced upflows
($V_{\rm D}>1.5\ {\rm km\ s}^{-1}$
with a size of
$2\arcsec\times 1\arcsec$
to $3\arcsec\times 2\arcsec$).
They seem to have
a slightly elongated structure
along the bridge.
The brightenings in AIA 1600 {\AA}
(contours)
are found
in the middle of the bridge.
Interestingly,
the brightenings are located
between the upflows of the convection cells:
see, e.g., the brightening
centered at $(x, y)=(17\arcsec, 7\arcsec)$
and surrounding blue-shifted regions.

These observed properties are
clearly seen in Figure \ref{fig:sp}(g),
which shows the averaged profiles
across the light bridge.
Here,
the vertical field $B_{z}$,
Doppler velocity $V_{\rm D}$,
field inclination from the vertical,
vertical current density $|j_{z}|$,
and the AIA 1600 {\AA} intensity
are plotted
against the $y$-axis.
The origin of the $y$-axis
is set at the middle
between the two maxima
of the current $|j_{z}|$.
In the light bridge
($|y|\lesssim 1\ {\rm Mm}$),
the vertical field shows a plateau
with relatively weaker,
almost horizontal field
($B_{z}\sim 0\ {\rm G}$,
$B_{x}\sim -1000\ {\rm G}$,
inclination $\sim 90^{\circ}$).
Therefore,
the positive field
seen in the light bridge structure
in the HMI magnetogram
(Figure \ref{fig:longterm}(f)
and Figure \ref{fig:tile}(j))
is mainly due to the projection effect
caused because the bridge is located
in the southwestern quadrant.
The Doppler velocity shows
an upflow ($V_{\rm D}\gtrsim 1.5\ {\rm km\ s}^{-1}$)
in the bridge.
At the same time,
it also has two local humps
around $y=\pm 0.8\ {\rm Mm}$,
which clearly shows
the existence
of multiple convection cells
in the bridge structure.
The electric current is confined
at the edges
of the bridge
($|j_{z}|\gtrsim 100\ {\rm mA\ m}^{-2}$
at $y\sim\pm 1.3\ {\rm Mm}$),
forming a clear boundary
between the bridge and the surrounding pores.
The downflows of the convection cells
are concentrated
just outside of
the bridge boundary:
see the slight dip
in the Doppler velocity
between $y=-1.2\ {\rm Mm}$
and $-1.8\ {\rm Mm}$
in this figure.
In the pores
($|y|\gtrsim 1.5\ {\rm Mm}$),
the magnetic field has
a large negative value
($B_{z}\sim -2000\ {\rm G}$)
and is more vertical,
while the Doppler velocity is much weaker
($V_{\rm D}\sim 0\ {\rm km\ s}^{-1}$).
The low-atmospheric brightening,
AIA 1600 {\AA},
peaks around $y=0.5\ {\rm Mm}$,
i.e., middle of the light bridge.
The peak of the brightening is
spatially offset
from the strong currents
at the both edges
and is located in between
the two convective upflows,
as is seen
in Figures \ref{fig:sp}(a) and (e).

\begin{figure}
  \begin{center}
    \includegraphics[width=80mm]{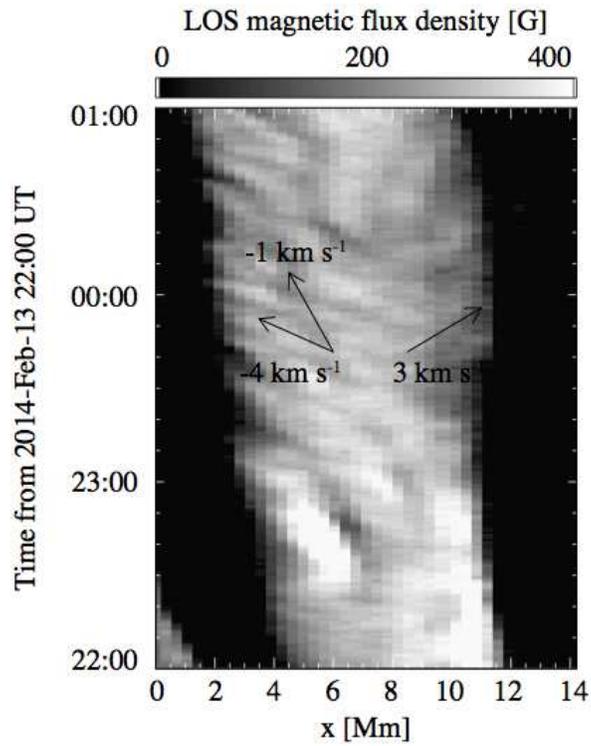}
  \end{center}
  \caption{Slit-time diagram
    of the light bridge
    of HMI LOS magnetogram.
    The slit is set
    along the $x$-axis,
    which corresponds
    to $y=6\farcs75$
    over the range
    of $10\arcsec\le x<30\arcsec$
    in Figures \ref{fig:sp}(a--f).
    Arrows indicate
    three different velocities.
  }
  \label{fig:lb_slit}
\end{figure}

The temporal evolution
of the light bridge
is given in Figure \ref{fig:lb_slit},
the time-sliced HMI magnetogram.
The slit of this diagram
is set along the $x$-axis,
which is roughly parallel to
the long axis
of the bridge.
In the eastern half
of Figure \ref{fig:lb_slit}
($x\lesssim 8\ {\rm Mm}$),
LOS magnetic fields show
a coherent apparent motion
to the east
with a typical pattern velocity
of $V_{x}=-1\ {\rm km\ s}^{-1}$
to $-4\ {\rm km\ s}^{-1}$.
In contrast,
the western half
($x\gtrsim 8\ {\rm Mm}$) shows
a westward propagation
with $V_{x}\sim 3\ {\rm km\ s}^{-1}$.
In Figure \ref{fig:lb_slit},
this divergent flow pattern
continues at least
for 2 hours
and the typical
time scale
of the pattern
is 10--15 minutes.

Combined with the Dopplergram
of Figure \ref{fig:sp}(e),
one can see that
the large-scale, long-term velocity structure
has an upflow
in the light bridge,
which turns into the divergent flow
in the horizontal direction
and finally sinks down
at the narrow edge
of the bridge.
Superposed on this large-scale flow pattern
are smaller-scale, shorter-lived convection cells
with an elongated shape.
The faint divergent magnetic pattern
in Figure \ref{fig:lb_slit} suggests
the continuous supply
of weak magnetic flux
from the solar interior,
transported by the large-scale upflow
of the light bridge.

\section{Dynamic Activity Phenomena
  \label{sec:results2}}

\subsection{Chromospheric Brightenings
  \label{subsec:brightenings}}

In Figure \ref{fig:tile},
we found that
a small-scale, intermittent brightening
located at the western end
of the light bridge
(brightening A)
repeatedly appear
in the chromospheric images
(AIA 1600 {\AA} and 1700 {\AA},
IRIS 1330 {\AA} and 1400 {\AA},
and SOT \ion{Ca}{2} H).
However,
in this figure,
similar brightenings
can also be found
outside of the bridge.
In order to approach
the cause of
such brightening events
in the chromosphere
(or the upper photosphere),
we then examine
the spatial distribution
and the spectral profiles
of the brightenings.

\begin{figure}
  \begin{center}
    \includegraphics[width=120mm]{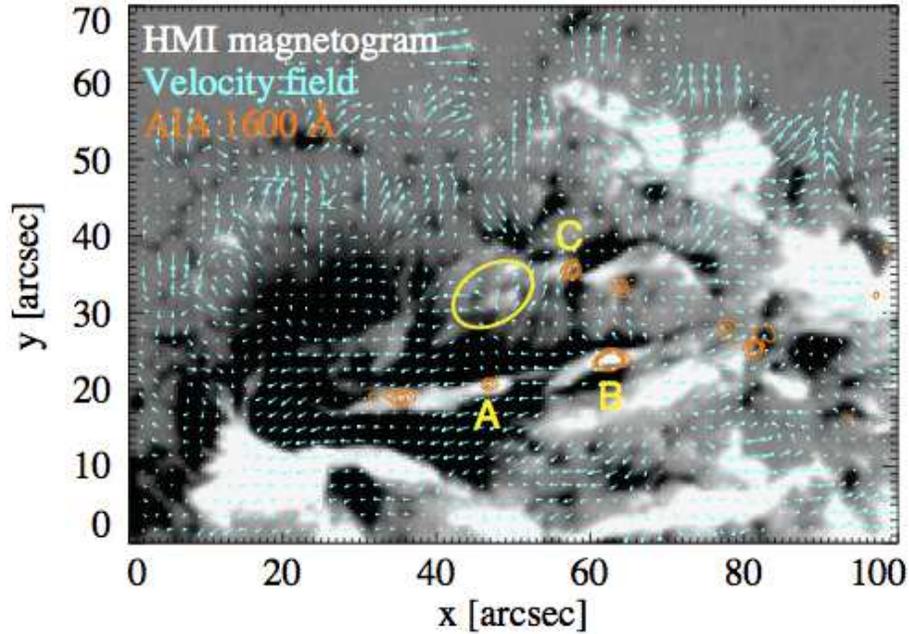}
  \end{center}
  \caption{Horizontal velocity field (arrows)
    calculated from the sequential HMI magnetograms
    for 1 hour from 00:00 UT, 2014 February 14,
    plotted over the magnetogram
    at 00:43 UT
    (gray scale saturating at $\pm 400\ {\rm G}$).
    Orange contours indicate
    the $10\sigma$, $15\sigma$, and $20\sigma$ levels
    above the mean
    of AIA 1600 {\AA} intensity
    at 00:43 UT.
    Three brightenings
    are denoted particularly
    by A, B, and C.
    Yellow ellipse shows
    one of the clear divergent regions,
    which indicates the flux emergence.
  }
  \label{fig:velocity}
\end{figure}

In Figure \ref{fig:velocity},
we show
the three intensity levels
of AIA 1600 {\AA} image
($10\sigma$, $15\sigma$, and $20\sigma$
above the mean
calculated from the entire $500\arcsec\times 250\arcsec$ FOV,
where $\sigma$ denotes
the standard deviation)
at 00:43 UT on 2014 February 14
(corresponding to Figure \ref{fig:tile}(h))
plotted on the HMI magnetogram
taken at the same time
(00:43 UT:
corresponding to Figure \ref{fig:tile}(j)).
One may see
from Figure \ref{fig:velocity} that
the brightening events
are, in many cases,
located at the mixed polarity regions,
where the positive and negative polarities
lie next to each other.
For example,
brightening A
is located on the western end
of the light bridge,
where the weak positive LOS magnetic field
is trapped between
the neighboring negative polarities
in the north and the south.
Similarly,
brightening B
is also on the elongated positive (LOS) field,
which is facing the negative fields
at the northern and southern edges.

Apart from these two events (A and B),
brightening C is located
in between the positive and negative polarities.
The arrows in Figure \ref{fig:velocity}
show the averaged horizontal velocity field
that is derived
from the sequential HMI magnetograms
between 00:00 UT and 01:00 UT,
2014 February 14,
using the local correlation tracking (LCT) method
\citep{nov88}.
Here,
the positive polarity
at the location of brightening C
is a new magnetic field
sourced from
a divergent region,
i.e., an emerging flux region
in the southeast
(indicated by yellow ellipse).
This newly emerging positive field
collides into the pre-existing negative field
and they cancel each other
at this location.
It should be noted here that
this local flux emergence
may also be responsible for
the merging of negative pores
surrounding the light bridge structure:
this flux emergence produces
the pore in the north
of the light bridge
and presses the pore
against the bridge
(see Figure \ref{fig:longterm}
and Section \ref{subsec:longterm}).

\begin{figure}
  \begin{center}
    \includegraphics[width=130mm]{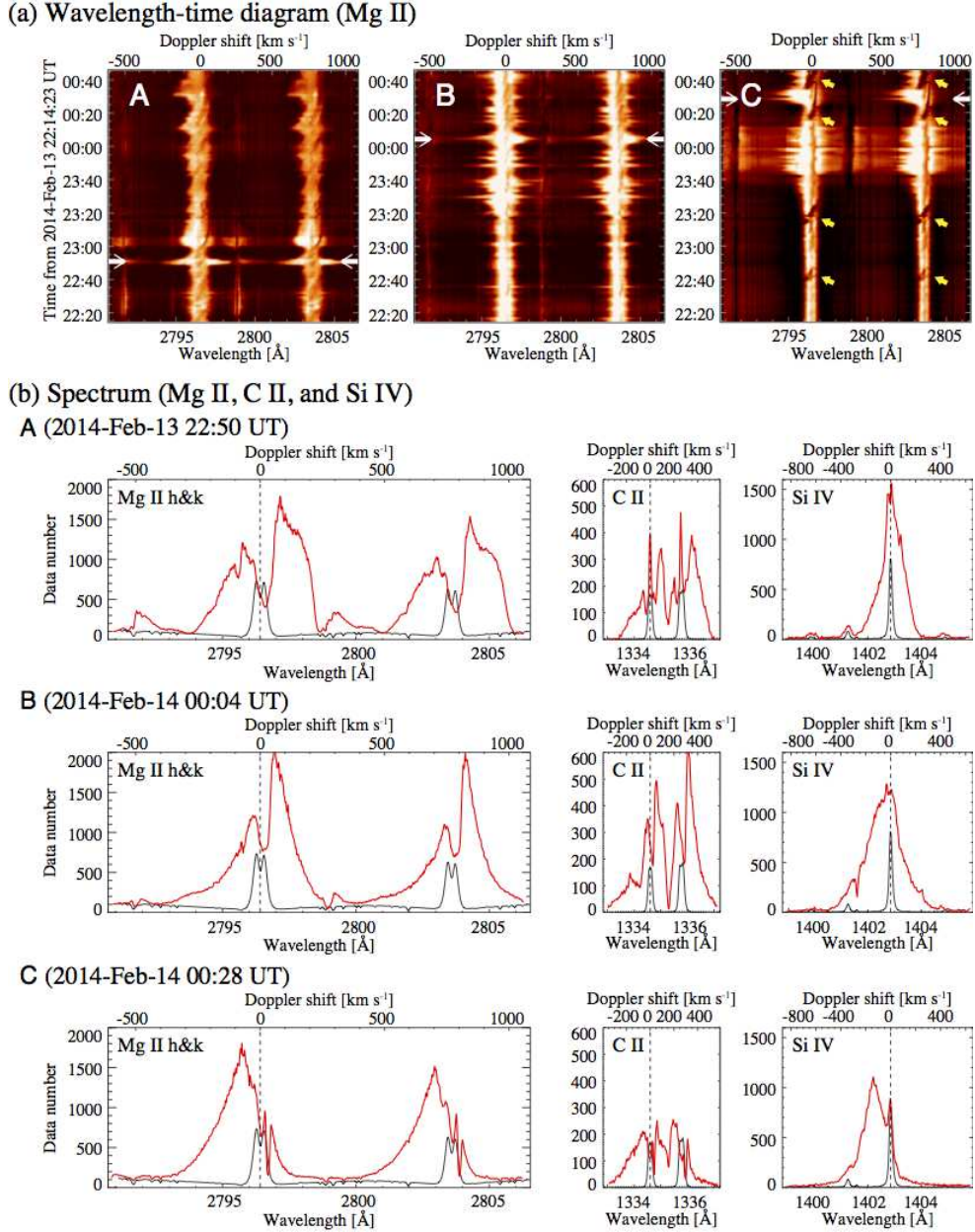}
  \end{center}
  \caption{IRIS spectra
    for the three brightening events.
    (a) Wavelength-time plots
    of \ion{Mg}{2} h \& k
    for brightenings A, B, and C.
    Each diagram shows a temporal evolution
    in a single spatial pixel
    of the IRIS raster scan.
    The top axis shows
    the Doppler shift
    relative to the rest wavelength
    of \ion{Mg}{2} k3.
    Yellow arrows indicate
    the dark transitions
    (see text for details).
    (b) Profiles of the three lines
    (\ion{Mg}{2}, \ion{C}{2}, and \ion{Si}{4})
    for the three events.
    Black lines show
    the averaged quiet-Sun spectra,
    while red lines
    show the profiles
    at the selected times
    indicated by white arrows in (a).
    Here,
    the quiet-Sun profiles
    of \ion{C}{2} and \ion{Si}{4}
    are multiplied by
    factors of 2 and 10,
    respectively.
  }
  \label{fig:iris}
\end{figure}

Figure \ref{fig:iris} shows
the IRIS spectra
for the three brightening events.
In the wavelength-time plots (panel (a)),
we can see that
each intensity enhancement occurs
repeatedly and frequently
at the same location.
The occurrence rate of the brightening is,
for example in case B,
once every few minutes.
Panel (b) shows
the spectra
at the time of brightenings
for three different lines
(\ion{Mg}{2}, \ion{C}{2}, and \ion{Si}{4})
and the referential averaged quiet-Sun spectra,
which are obtained
from the less active area
within the FOV
of the same IRIS observation.
The central reversal
in the \ion{Mg}{2} reference profiles
is the self-absorption
due to large opacity.
Here,
the spectra at A and B
are intensified
compared to the quiet-Sun profiles
and are broadened
toward both red and blue wings.
These higher amplitude and broadened profiles
are similar
to the recent IRIS observation
of brightening events
at the flux cancellation sites
in an emerging AR
\citep{pet14}
and the Ellerman bombs
\citep{vis15}.
Such spectral profiles are,
in many cases,
interpreted as
the local heating of
chromospheric plasma
via magnetic reconnection
and the bi-directional outflow
from the reconnection region
\citep[e.g.,][]{pet14}.
Therefore,
it is also possible
that the enhanced and broadened profiles
seen in Figure \ref{fig:iris}
represents
the energy releasing
by magnetic reconnection.
This interpretation is further supported
by the spatial distribution
of the brightening events.
In Figure \ref{fig:velocity},
the brightenings
were found to exist
in the mixed polarity regions
and some of them showed
flux cancellation,
which also indicates that
the brightenings are caused
by magnetic reconnection.

In contrast,
IRIS spectral profiles of
brightening C
in Figure \ref{fig:iris}(b)
look different
from those of A and B.
For instance,
the \ion{Mg}{2} profile
of brightening C
is seemingly single-peaked
showing a strong upflow
($\sim -70\ {\rm km\ s}^{-1}$:
blue-shifted)
rather than the double-peaked profile
of events A and B
with clear central reversals.
Upon closer inspection,
however,
one finds that
the spectrum
does have a central reversal
with another dip
on the red side of the reversal.
Here, the Doppler shift of the dip
from the rest wavelength
of the line center
is $+32.7\ {\rm km\ s}^{-1}$ (red-shifted),
while the central reversal
is located at $+5.5\ {\rm km\ s}^{-1}$ (red-shifted).
A similar dip is also seen
in the \ion{C}{2} spectrum.
This anomalous dip
corresponds to
the dark feature
(intensity reduction)
in the wavelength-time plot
(marked by yellow arrows in panel (a)).
This feature appears repeatedly
with a lifetime of 10--20 minutes
(four events: 22:39--22:47 UT and 23:13--23:25 UT
on February 13
and 00:16--00:38 UT and from 00:35 UT
on February 14),
each time showing
a sudden appearance
of the blue shift and
a smooth transition
from the blue to the red side.
We will discuss
the cause of this anomalous intensity reduction
in Section \ref{subsec:surges}.

\subsection{Ejection of Dark Surges
  \label{subsec:surges}}

\begin{figure}
  \begin{center}
    \includegraphics[width=160mm]{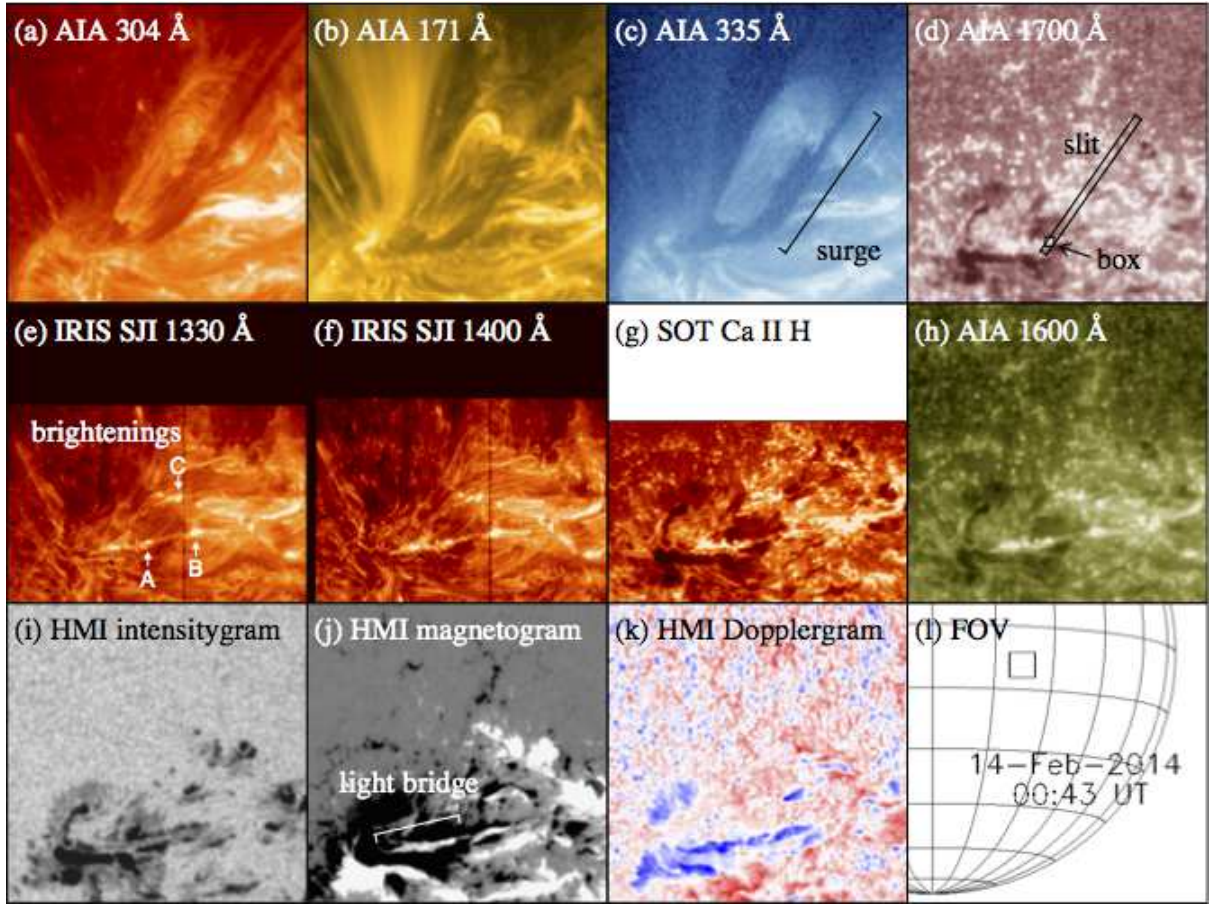}
  \end{center}
  \caption{Composite of (a--d and h) AIA images,
    (e and f) IRIS SJIs,
    (g) SOT Ca image,
    and (i--k) HMI intensitygram,
    LOS magnetogram (saturating at $\pm 400\ {\rm G}$),
    and Dopplergram (saturating at $\pm 2\ {\rm km\ s}^{-1}$),
    all taken at around 00:43 UT, 2014 February 14.
    Panel (l) shows the location of the FOV
    (each panel has a $100\arcsec\times 100\arcsec$ FOV)
    and the time of the observation.
    Three continuous brightenings
    in the chromospheric images
    are marked
    as A, B, and C
    in panel (e),
    the dark surge is indicated
    in panel (c),
    and the light bridge
    is shown in panel (j).
    In panel (d),
    we show the locations
    of the slit and the box,
    which are used in the analysis
    in Section \ref{subsec:surges}.
  }
  \label{fig:tile}
\end{figure}

\begin{figure}
  \begin{center}
    \includegraphics[width=100mm]{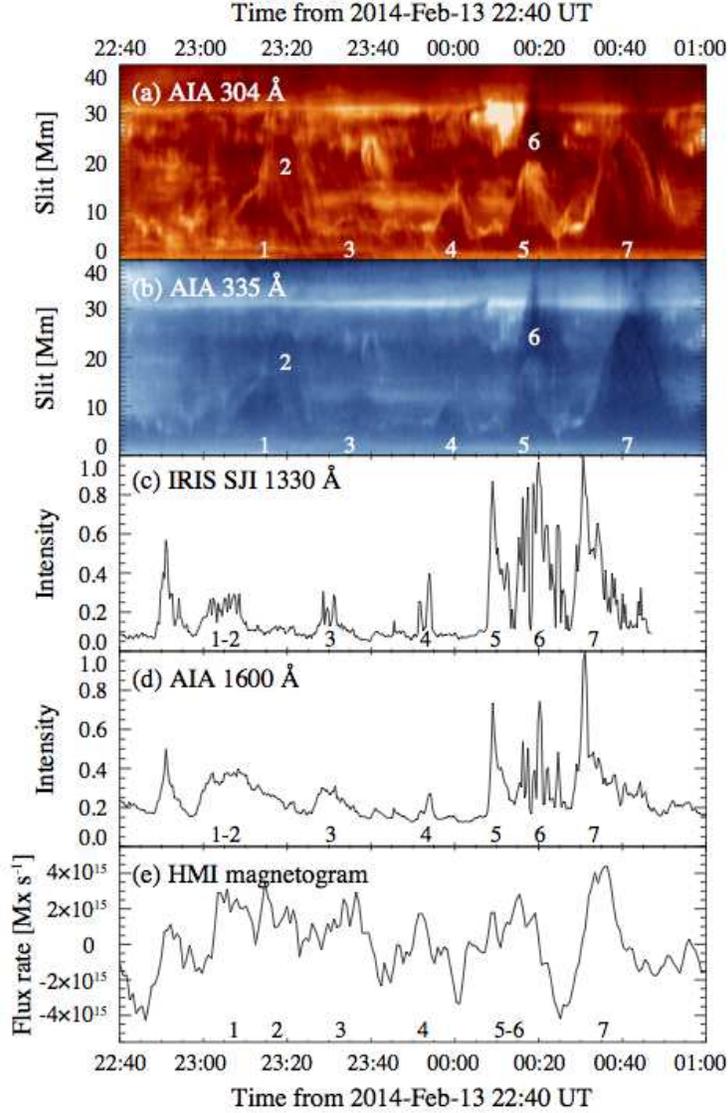}
  \end{center}
  \caption{Slit-time evolution
    for (a) AIA 304 {\AA}
    and (b) AIA 335 {\AA} images,
    normalized lightcurves
    for (c) IRIS SJI 1330 {\AA}
    and (d) AIA 1600 {\AA},
    and (e) the (smoothed) flux ``decay'' rate
    $-d\Phi/dt$
    measured using HMI magnetogram
    (see text for details).
    The slit used in (a) and (b),
    which is shown by a rectangle
    in Figure \ref{fig:tile}(d),
    has a width of $2\farcs 5$
    and the AIA intensities
    are averaged over the width.
    The lightcurves in (c) and (d)
    and the flux rate in (e)
    are measured
    within the box shown also
    in Figure \ref{fig:tile}(d).
    The size of the box is
    $3\arcsec\times 3\arcsec$.
    In each panel,
    we numbered 7 events
    to provide better correspondence.
  }
  \label{fig:slit}
\end{figure}

In the EUV images
of Figure \ref{fig:tile},
we found
the recurrent ejection
of a dark surge (jet)
from the light bridge,
especially from brightening A.
In order to investigate
the nature of the dark surges,
along with its relationship
with the chromospheric
brightenings
and the magnetic field
of the light bridge,
we examine the temporal evolutions
of the surge,
the brightening,
and the surface magnetic field.
In Figures \ref{fig:slit} (a) and (b),
we plot the time-slit diagrams
for the two EUV images
(AIA 304 {\AA} and 335 {\AA}),
where the slit is set
along the surges
as indicated
in Figure \ref{fig:tile}(d).
One can see
from Figures \ref{fig:slit}(a) and (b) that
there is a recurrent extension
of dark features
(i.e., surges)
with a typical lifetime
of 10--20 minutes.
The observed surges
have a parabolic trajectory,
which indicates that
the ejected material is
impulsively accelerated upward
and decelerated by gravity.
In all EUV channels,
the surges are seen in absorption,
implying that
the material is cool and dense.
However,
in some channels
such as AIA 304 {\AA}
in Figure \ref{fig:slit}(a),
the front of each surge
(the rim of the parabolic profile)
appears in emission.

Table \ref{tab:surges} summarizes
the properties of the surges
(lifetime, length, and acceleration).
The length parameter
in this table
is defined as
the maximum spatial extension
of each surge
along the slit
in Figure \ref{fig:slit}(b),
which is not necessary the same as
the maximum height
of the surge
because of the projection effect
and the inclination of the surge.
Also,
we calculated
the acceleration of the surge
from the length and lifetime
as $a_{\rm DS}=8L_{\rm DS}/t_{\rm DS}^{2}$,
where $a_{\rm DS}$, $L_{\rm DS}$, and $t_{\rm DS}$ are
the acceleration, length,
and lifetime
of the dark surge,
respectively,
under the assumption that
the launched plasma
is decelerating
at a constant rate.
The obtained accelerations
($1.1\times 10^{4}\ {\rm cm\ s}^{-2}$
to $4.0\times 10^{4}\ {\rm cm\ s}^{-2}$)
are of the same order of magnitude
as the surface gravity
($2.7\times 10^{4}\ {\rm cm\ s}^{-2}$)
indicating the gravitational deceleration.

In Figures \ref{fig:slit}(c) and (d),
we plot the normalized lightcurves
at the brightening A
for the two UV images,
IRIS 1330 {\AA} SJI
and AIA 1600 {\AA},
respectively.
The lightcurves are
measured within the box
of the size
of $3\arcsec\times 3\arcsec$
indicated in Figure \ref{fig:tile}(d).
Here,
the brightening events
have a duration
of 10--20 minutes
and, at the same time,
the lightcurves show
much more rapid fluctuations
of a time scale over 3--5 minutes.
Compared to the EUV images
(Figures \ref{fig:slit}(a) and (b)),
it is remarkable that
each intensity enhancement
in the lightcurves occurs
just prior to the dark surge ejection
(compare the numbered events
in these panels
of Figure \ref{fig:slit}).
In each event,
the lightcurve reaches its maximum
about 5--15 minutes
before the surge attains
its maximum length.
This temporal relationship
suggests the causality
between the
chromospheric
brightenings
and the dark surges.

Table \ref{tab:surges} lists
the duration and the maximum normalized intensity
of each brightening event
measured from the AIA 1600 {\AA} lightcurve
(Figure \ref{fig:slit}(d)).
For evaluating these values,
we do not use
IRIS SJI data (panel (c))
since SJI contains the slit
and some contamination
(see Figures \ref{fig:tile}(e) and (f))
and, in case of strong brightenings,
some saturated pixels.
These factors may be reflected
in the IRIS SJI lightcurve.

\begin{figure}
  \begin{center}
    \includegraphics[width=120mm]{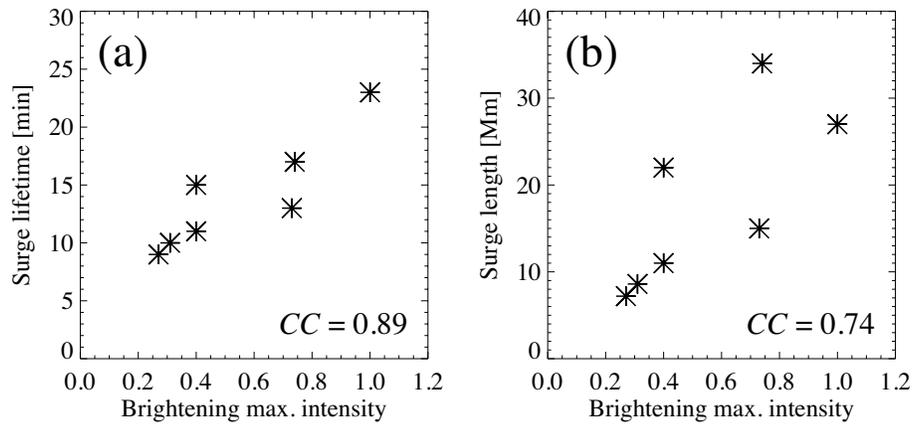}
  \end{center}
  \caption{Correlations
    among brightening and surge parameters
    shown in Table \ref{tab:surges}.
    Correlation coefficient
    (linear Pearson correlation coefficient, $CC$)
    is indicated
    in the bottom right
    of each panel.
  }
  \label{fig:slit_cc}
\end{figure}

The relationship between 
the two dynamic phenomena
in the light bridge,
i.e., the brightenings and the surges,
can be found in Figure \ref{fig:slit_cc}.
This figure shows
the correlations
among parameters
in Table \ref{tab:surges},
the maximum normalized intensity
of the brightenings
and the lifetime and length
of the surges.
They show good correlations:
the corresponding correlation coefficients
(linear Pearson correlation coefficient, $CC$)
for panels (a) and (b)
are $CC=0.89$ and $0.74$,
respectively.
That is,
the length and lifetime
of the surges
become larger
with the brightening intensity.
Although we have only a sample of 7 event pairs,
these good correlations
clearly point to the physical relationship
between the brightenings
and the dark surges.

In Section \ref{subsec:brightenings},
based on the spatial distribution
of the brightening events
and their IRIS spectral profiles,
we interpreted
the cause
of the brightenings
as magnetic reconnection.
Figure \ref{fig:slit}(e)
may provide
further support
to this interpretation.
This diagram shows
the flux ``decay'' rate
$-d\Phi/dt$,
where $\Phi$ is
the unsigned total magnetic flux,
\begin{eqnarray}
  \Phi = \int_{S} |B_{\rm LOS}|\,ds,
\end{eqnarray}
$B_{\rm LOS}$ is
the LOS magnetic flux density
in the HMI magnetogram,
and $S\, (=3\arcsec\times 3\arcsec)$
is an area of the box
in which the brightening lightcurves
(i.e., Figures \ref{fig:slit}(c) and (d))
are measured.
This decay rate is positive
when the total magnetic flux decreases.
In Figure \ref{fig:slit}(e),
it is clearly seen that
the flux rate becomes positive
around the timing of brightening events.
In contrast,
when there is no prominent brightening,
the decay
rate remains negative.
This trend indicates that
the magnetic flux
is continuously supplied
to this region
and is consumed during the brightenings,
which we think is
the magnetic reconnection
(flux cancellation).
As a result of the repeated reconnection,
the plasma is ejected upward
to create the recurrent dark surges.

The IRIS wavelength-time diagram of \ion{Mg}{2}
(Figure \ref{fig:iris}(a))
at location C
shows smooth transitions
of intensity reduction
from blueshift to redshift
relative to the rest wavelength
of the line core
(see Section \ref{subsec:brightenings}).
By comparing this figure
with Figures \ref{fig:slit}(a) and (b),
one may notice the temporal correspondence
between the intensity reductions
and the dark surges.
For example,
the three reductions
starting at 23:13 UT on February 13
and 00:16 UT and 00:35 UT on February 14
occur almost
at the same times
as the surges numbered 2, 6, and 7.
Also,
both types of features
(the intensity reduction
and the dark surge)
persist for 10--20 minutes.
In addition to that,
in the EUV time-slit diagrams
of Figures \ref{fig:slit}(a) and (b),
these three surges (2, 6, and 7)
are the only events
that go beyond
the location of the brightening C,
which is around
15--17 Mm on the slit
but is not seen
in the EUV channels.
Therefore,
from the temporal and spatial correspondence,
we can expect that
the repeated intensity reductions
in the IRIS spectrum
of brightening C
to be
caused by the recurrent dark surges
that passed over
and occulted
the brightening
along the IRIS LOS.
The smooth transition
from the blueshift to the redshift
may indicate that
the surge changes
from the ascending to descending motion.
We can furthermore see that
the dark surges
reach
chromospheric temperatures,
e.g., 10,000 K for \ion{Mg}{2}
\citep{dep14}.

\section{Summary of Observational Results
  \label{sec:summary}}

In the previous sections,
we analyzed the light bridge structure
in the developing AR 11974
and the activity phenomena
around the bridge.
The light bridge had
a size of $\sim 22\ {\rm Mm}\times 3\ {\rm Mm}$
with a lifetime of $\sim 2$ days
and was surrounded
by merging negative
polarity pores.
Images in the
AIA 1600 {\AA} and 1700 {\AA},
IRIS 1330 {\AA} and 1400 {\AA} SJI,
SOT \ion{Ca}{2} H channels
showed
small-scale intermittent brightenings,
while the EUV data
(AIA 304 {\AA}, 171 {\AA}, 335 {\AA}, etc.)
revealed the recurrent ejection
of dark surges.
By conducting the series of detailed analysis,
we found the following:

\begin{itemize}
\item Photospheric magnetic field
in the light bridge
was relatively weaker
($\sim 1000\ {\rm G}$)
and almost horizontal,
while the surrounding pores showed
strong
($\sim 2000\ {\rm G}$),
almost vertical fields
of the negative polarity.
Because of the magnetic shear
between the bridge and the pores,
the boundary layer
had a strong vertical current
($\gtrsim 100\ {\rm mA\ m}^{-2}$)
in the photosphere,
which is a favorable site
for magnetic reconnection.
The vertical motion
in the light bridge
was basically upward
($\gtrsim 1\ {\rm km\ s}^{-1}$),
while the narrow edges
showed strong downflows
(down to $-6\ {\rm km\ s}^{-1}$).
At the same time,
the bridge showed the existence
of multiple smaller-sized convection cells inside
($>1.5\ {\rm km\ s}^{-1}$;
$\sim 2\arcsec\times 1\arcsec$
to $3\arcsec\times 2\arcsec$).
It was found that
the chromospheric (or upper-photospheric)
brightenings
were located
in the middle of the bridge,
i.e., deviated from the location
of photospheric currents.
Also,
the brightenings were seen
in between the small-scale upflows.
These results show that
the large-scale, long-term velocity structure
of the light bridge is
driven by an upflow,
which then turns into
the horizontal divergent motion
($V_{x}$ being $\pm$ a few km s$^{-1}$)
and finally sinks down
at the narrow edge
of the bridge.
The faint divergent pattern
of magnetogram
indicates the existence
of continuous supply of magnetic flux
by the large-scale upflow.
The typical timescale
of the propagation pattern
was 10--15 minutes.

\item The
chromospheric
brightenings
occurred in the mixed polarity regions,
some of which showed
flux cancellation.
For example,
brightening A was
on the western end
of the light bridge
of weak positive LOS fields,
which was facing
the surrounding negative pores.
Brightening B had
a similar magnetic context.
On the other hand,
brightening C was located
at the flux cancellation site
of the newly-emerging positive field
and the pre-existing negative field.
IRIS spectra of the brightenings
showed an enhanced and broadened profiles
akin to those associated
with Ellerman bombs.
From these results,
we interpret the brightenings
as the local heating
of chromospheric plasma
by magnetic reconnection.
Broadening of the IRIS spectra
may indicate the bi-directional outflow
from the reconnection region.

\item The brightenings
preceded the recurrent dark surges
observed in the EUV images
by 5--15 minutes.
The surges were ejected
from the light bridge,
especially from brightening A.
The typical lifetime of the surges
was 10--20 minutes.
The surges appeared
basically in absorption,
indicating that the ejected material
is cool and dense.
However,
in some channels such as AIA 304 {\AA},
the front edges
of the surges
were seen in emission.
The surges showed
a parabolic profile
in time-distance diagrams,
implying the existence of
gravitational deceleration.
In fact,
the accelerations
measured for the surge events
were of the same order of magnitude
as the surface gravity.
The typical duration
of the brightenings
was 10--20 minutes,
while at the same time
they showed a rapid fluctuation
with a period
of 3--5 minutes.
It was also found that
the length and lifetime
of the surges
are well correlated with
the brightening intensity.
In addition to that,
in the photosphere
below brightening A,
we found the decrease of
unsigned magnetic flux
during the brightening events
and the continuous supply of the flux
during the rest
of the period.
Furthermore,
the surges were observed
in IRIS \ion{Mg}{2} and \ion{C}{2} spectra
as temporal intensity reductions
with a smooth transition
from the blueshift to the redshift.
The above-mentioned results indicate that,
along with the heating of the plasma
that can be observed
as the brightenings,
the released energy
of the magnetic reconnection
was converted also
into the kinetic energy
by launching
the cool, dense
chromospheric
plasma
into the higher atmosphere
in the form of dark surges.
\end{itemize}

\section{Discussion
  \label{sec:discussion}}

\subsection{General Picture}

Figure \ref{fig:illust}(a) illustrates
the general picture
that explains
the activity phenomena
in the analyzed AR.
Here,
one of the most prominent events
is the repeated and intermittent flickering
in the chromospheric images.
These brightenings are
the local heating
of chromospheric
(or upper-photospheric)
plasma
via magnetic reconnection
and that is why
they are seen mainly
in the mixed polarity regions.
Brightening A occurs
in the light bridge,
which is the weak,
almost horizontal field
trapped between the negative pores
of almost vertical field.
It shows a repeated intensity enhancement
with a typical duration
of 10--20 minutes.
Within each event,
it also has a rapid fluctuation
with a timescale of 3--5 minutes.
Brightening B has
a similar magnetic background,
while brightening C
is caused by the flux cancellation
of a pre-existing negative field
and a positive field
from the emerging flux region.
Because of this flux emergence,
the negative pores
around the light bridge
come closer to merge with each other.
Eventually the pores
form a sunspot,
leading to the vanishing
of the light bridge.

As well as the local heating,
a part of the released energy
of magnetic reconnection
is converted also
into the kinetic energy
by launching
cool and dense chromospheric plasma
into the higher atmosphere
in the form of dark surge,
which is observed
as absorption
in EUV images.
The surges are recurrently ejected
with a typical lifetime of 10--20 minutes,
each following the brightening.
Due to gravity,
the launched cool plasma is pulled back
to the surface,
showing a parabolic trajectory
in the EUV time-slices
and a smooth transition
from the blueshift to the redshift
in the IRIS spectra.

\subsection{Light Bridge
  and its Brightenings}

\begin{figure}
  \begin{center}
    \includegraphics[width=75mm]{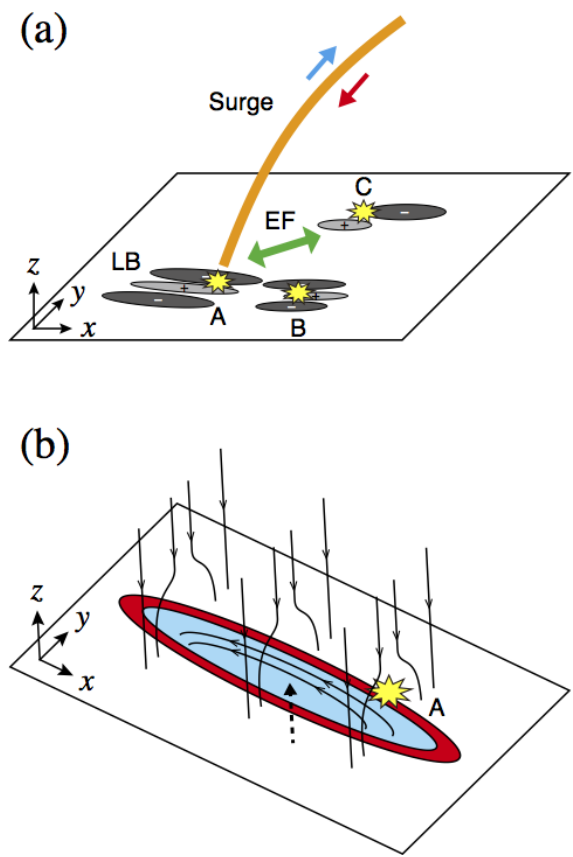}
  \end{center}
  \caption{
    {\footnotesize
    (a) Illustration
    summarizing activity phenomena
    in AR 11974.
    The chromospheric brightenings
    are found in mixed polarity regions,
    where the positive field
    (lighter ellipse with ``$+$'' sign)
    and negative field
    (darker ellipse with ``$-$'' sign)
    lie next to each other.
    Brightening A is located
    at the light bridge (LB)
    and ejects dark surges
    (thick orange line)
    into the higher atmosphere,
    which shows a shift
    from the upward (blue arrow)
    to downward (red arrow) motion.
    The light bridge becomes narrower
    because of the neighboring emerging flux
    (EF: green arrow).
    Brightening B has
    a similar magnetic structure
    as brightening A.
    Brightening C is caused by
    flux cancellation
    between the positive field
    sourced from the emerging flux (EF)
    and the negative polarity.
    Coordinate system $(x, y, z)$
    is consistent with
    that we used
    in the analysis
    of this paper.
    (b) Schematic illustration
    of the light bridge.
    The light bridge shows
    a broad blue-shifted region (blue)
    indicating the large-scale upflow
    from the solar interior
    (dashed arrow).
    The upflow carries
    the horizontal magnetic flux
    (solid lines with arrows in the bridge),
    which reconnects with the vertical flux
    of the surrounding pores
    (solid lines with arrows outside the bridge)
    and shows brightening events.
    One representative event,
    brightening A,
    is located in the western edge
    (denoted as A).
    The flow drains down
    in the narrow downflow lane
    (red).
    }
  }
  \label{fig:illust}
\end{figure}

Figure \ref{fig:illust}(b) shows
a schematic illustration
of the light bridge.
The bridge has
a large-scale, long-lasting upflow,
which turns into a diverging flow
in the horizontal direction.
The flow sinks down
in the narrow lane
at the edge
of the bridge.
Transported by the large-scale upflow,
magnetic flux
is continuously provided
from the solar interior
to the light bridge
as a relatively weaker, horizontal field,
which is in agreement with previous observations
\citep{bec69,lit91,rue95,lek97,jur06}.
At the same time,
the bridge contains
small-scale (arcsec-sized)
local convection cells
with lifetimes
of order of 10 minutes.
The delivered magnetic flux
from the interior
has a pattern of 10--15 minutes,
probably due to the local convection.
As a result,
magnetic reconnection
takes place repeatedly
in the light bridge region
between weaker horizontal fields
of the light bridge
and vertical fields
of the surrounding pores
with an observed period
of 10--20 minutes.
The reconnection is seen
as the brightenings,
while the ejected outflow
reaches coronal heights
as the dark surges.

One of the brightenings
that are observed
in this light bridge,
namely brightening A,
is located at the western end
of the bridge.
The reason for
the repeated brightenings
at this location
may be that
the footpoints
of the transported horizontal fields
are more likely to be positive
in the western part
of the bridge,
since the observed horizontal fields
are mainly eastward
(see brightening A
in Figure \ref{fig:illust}(b)).
Therefore,
as the transported fields
pile up at the western end,
the magnetic shear
between the positive fields
within the light bridge
and the vertical negative fields
of the surrounding pores
increases.
And thus,
the western end
becomes a favorable site
for the magnetic reconnection,
which is observed as
series of brightenings.

By comparing
the photospheric vertical currents
and the
chromospheric
brightenings
in Figure \ref{fig:sp},
we noticed
that the brightenings are located
in the middle of the light bridge,
while the current sheets are
at the edge of the bridge.
One possible explanation
for this difference
is the canopy structure
of the magnetic fields
above the light bridge
\citep{lek97,jur06}.
Because of the intrusion
of the horizontal field
of the light bridge,
surrounding vertical fields
of the pores
need to fan out
over the bridge
until they meet the field lines
from the opposite side
(see the vertical fields
in Figure \ref{fig:illust}(b)).
In the higher altitude,
or in the formation layer
of the UV lines,
the width of the horizontal field region
is narrower and more confined
to the middle
of the bridge.
And thus,
the UV brightenings
are seen in the bridge center,
which is deviated
from the location
of the photospheric current sheet
at the both edges.
The observed magnitude
of the vertical current density,
$|j_{z}|\gtrsim 100\ {\rm mA\ m}^{-2}$,
is consistent with
previously reported values
\citep[e.g.,][]{jur06}.
Similarly to our observational results,
\citet{shi09} detected
the current density
higher than $100\ {\rm mA\ m}^{-2}$
along the light bridge.
They thought that
the strong current was responsible
for the intermittent chromospheric jets
observed in \ion{Ca}{2} H images.

In Figure \ref{fig:sp},
we found the existence
of multiple smaller-sized convection cells
in the light bridge,
which is consistent with previous observations.
\citet{ber03} reported
the central dark lanes
in the light bridge,
indicating the multiple convection cells
\citep[see also][]{sob94,hir02,lag14}.
In our case,
small-scale cells had
elliptic shapes
with elongation
along the direction
of the light bridge.
This
may be the consequence
of the horizontal field
parallel to
the light bridge
\citep[see review of simulations by][]{che14}.

\subsection{Physical Mechanism
  of the Dark Surges}

The chromospheric
brightenings
that were repeatedly observed
in the UV data
were followed by
the elongations
of dark absorption features
in the EUV images,
which we call dark surges.
Also,
the surges were composed of
cool, dense (i.e. chromospheric) plasma
with a temperature
of down to 10,000 K.
The observed dark surge
shows a very similar behavior
to H$\alpha$ surges
\citep{roy73}.
According to \citet{asa01},
the ejections from
the light bridge
in H$\alpha$ and 171 {\AA} images
showed that
the H$\alpha$ surges
are ejected intermittently and recurrently
from the light bridge
and that
the velocities and timings
of the 171 {\AA} ejections
are the same as
those of the H$\alpha$ surges.
\citet{roy73} reported that
all surges observed
extend from Ellerman bombs.

What is the physical mechanism
that launches
the cool chromospheric material
into the higher atmosphere
from the light bridge?
One possible answer
is the sling-shot effect
of magnetic reconnection.
\citet{yok95,yok96} conducted
a numerical simulation
of the reconnection
between the emerging magnetic field
and the pre-existing coronal field.
In their model,
cool chromospheric plasma
is ejected upward
by the tension force
of curved reconnected field
(sling-shot effect),
which may be observed
as H$\alpha$ surges.
Considering the similarities
between the H$\alpha$ surges
and the dark surges
in our analysis,
we can speculate that
the dark surges
are basically produced by
the same physical mechanism.

However,
it is still difficult
to launch such a cool, dense material
to the higher altitude.
If we assume that
the magnetic energy
of the light bridge,
$B_{\rm LB}^{2}/(8\pi)$,
is completely converted
into the potential energy
of the dark surge,
$\rho g H_{\rm DS}$,
the height of the surge
is given as
$H_{\rm DS}=B_{\rm LB}^{2}/(8\pi\rho g)$,
where $B_{\rm LB}$, $\rho$, $g$, and $H_{\rm DS}$
are the field strength of the light bridge,
the plasma density,
the gravitational acceleration,
and the height of the dark surge,
respectively.
If we adopt
$B_{\rm LB}\sim 2,000\ {\rm G}$,
$\rho\sim 1\times 10^{-8}\ {\rm g\ cm}^{-3}$
(approximate density around the temperature minimum),
and $g\sim 2.7\times 10^{4}\ {\rm cm\ s}^{-2}$,
then we obtain
$H_{\rm DS}\sim 6\ {\rm Mm}$.
However,
the apparent lengths
of the surges
in our observation
were up to 34 Mm
(see Table \ref{tab:surges}),
which indicates that
the surge height, $H_{\rm DS}$,
is a few 10 Mm.
By assuming
higher reconnection altitudes
and thus applying smaller $\rho$,
we may obtain
larger $H_{\rm DS}$.
However,
at the same time,
$B_{\rm LB}$ may also become smaller,
which may results in
insufficient surge heights $H_{\rm DS}$.

One promising idea
to overcome this discrepancy
is the acceleration
by a slow-mode shock
\citep[e.g.,][]{shi07}.
\citet{tak13} conducted numerical simulations
similar to those of \citet{yok96}
but of the case that
the magnetic reconnection
takes place
below the transition region,
say, in the chromosphere.
In such a case,
a slow shock is produced
and interacts with the transition region.
As the transition region
is lifted up by the slow shock,
the chromospheric plasma
extends into much higher altitude.
Therefore,
the observed dark surge
in this study
can be launched
by the reconnection
with the help
of slow shock.
Moreover,
the lifting
of the transition region
can explain
the emission
in the front edge
of the dark surges
seen in some AIA channels,
e.g., 304 {\AA}
as in Figure \ref{fig:slit}(a),
which are sensitive to
the transition region plasma.

\section{Conclusion and Outstanding Questions
  \label{sec:conclusion}}

In this study,
we analyzed
the light bridge structure
in a developing AR
and dynamic activity phenomena
related to the light bridge,
i.e., chromospheric brightenings
and dark surges.
The observations lead us to conclude
that the brightenings and dark surges
are driven
by magnetoconvective evolution
within the light bridge
and its interaction
with the umbral surroundings.
Namely,
the convective upflow
in the bridge
continuously transports
the horizontal fields
to the surface,
which reconnects with adjacent vertical fields,
resulting in
the repeated and intermittent
brightenings and surge ejections.
However,
we still have
some remaining problems.

First of all,
the creation of
the light bridge structure
remains unclear.
In the target AR
of this study,
the light bridge appeared
between the two merging pores
of the negative polarity.
Although the formation process
of a light bridge
in a mature sunspot
was observationally studied
in detail
by \citet{kat07},
the light bridge formation
in a developing sunspot
needs further investigation.

The second issue is
the large-scale flow structure
in the light bridge.
In the present study,
the Dopplergrams revealed
the existence of
a large-scale upflow
within the bridge.
This upflow transports
the magnetic flux
from the convection zone
to the surface layer
and causes magnetic reconnection
with the surrounding pore fields,
resulting in the production of
dynamic events.
However,
the driver of the large-scale upflow
and its transportation mechanism
of magnetic flux
were not revealed
in the present study.

The fine-scale structure
of the light bridge
is another important topic.
For example,
the present observation
as well as previous observations
\citep[e.g.,][]{ber03}
revealed the existence
of local convection cells
inside the light bridge.
However,
it is difficult
to resolve the such arcsec-sized cells
or even subarcsec-sized downflow lanes.
Also,
it was found that
the electric current,
$|j_{z}|$,
is confined
to the narrow boundary layer
between the bridge and the external pores.
Therefore,
revealing the detailed structure
of the total current,
$|\mbox{\boldmath $j$}|$
instead of $|j_{z}|$,
may contribute
to better understanding
of the occurrence
of magnetic reconnection.

These issues will be addressed
in \citetalias{tor15b}
of this series.
The first two topics,
the formation of the light bridge
in a developing AR
and its large-scale flow structure,
are necessarily linked to
the dynamical evolution
of an emerging flux region.
Also,
we need fine-scale and three-dimensional analysis
on the bridge structure
to reveal the nature
of local convection
and the structure
of total current.
Therefore,
in \citetalias{tor15b},
we will analyze
the numerical simulation data
of the large-scale three-dimensional flux emergence
including thermal convection
by \citet{che10}
and investigate
the light bridge structure
in a spot formation region.



\acknowledgments

The authors thank
the anonymous referee
for improving the manuscript.
Data are courtesy
of the science teams
of {\it Hinode}, {\it SDO} and IRIS.
{\it Hinode} is a Japanese mission
developed and launched by ISAS/JAXA,
with NAOJ as domestic partner
and NASA and STFC (UK)
as international partners.
It is operated by these agencies
in co-operation with ESA and NSC (Norway).
HMI and AIA are instruments
on board {\it SDO},
a mission for NASA's
Living With a Star program.
IRIS is
a NASA small explorer mission
developed and operated
by LMSAL
with mission operations
executed at NASA Ames Research center
and major contributions
to downlink communications
funded by ESA
and the Norwegian Space Centre.
This work was carried out
on the Solar Data Analysis System
operated by the Astronomy Data Center
in cooperation with
the Hinode Science Center
of NAOJ.
This work was supported
by JSPS KAKENHI
Grant Number 26887046 (PI: S. Toriumi)
and 25220703 (PI: S. Tsuneta). 
MCMC acknowledges
support by NASA contracts NNG09FA40C (IRIS),
NNG04EA00C ({\it SDO}/AIA)
and NNM07AA01C ({\it Hinode}/SOT),
and grant NNX14AI14G
(Heliophysics Grand Challenges Research).

\clearpage

\clearpage

\clearpage

\clearpage

\clearpage

\clearpage

\clearpage

\clearpage

\clearpage

\clearpage

\begin{deluxetable}{lccccccc}
  \tabletypesize{\scriptsize}
  \tablecaption{Properties
    of the Dark Surge
    and Brightening\label{tab:surges}}
  \tablewidth{0pt}
  \tablehead{
    \colhead{\#} & \colhead{} &
    \multicolumn{3}{c}{Dark surge\tablenotemark{a}} &
    \colhead{} &
    \multicolumn{2}{c}{Brightening\tablenotemark{b}}\\
    \cline{3-5} \cline{7-8}
    \colhead{} & \colhead{} &
    \colhead{Lifetime} & \colhead{Length} &
    \colhead{Acceleration} &
    \colhead{} &
    \colhead{Duration} & \colhead{Max. intensity}\\
    \colhead{} & \colhead{} &
    \colhead{(min)} & \colhead{(Mm)} &
    \colhead{($10^{4}\ {\rm cm\ s}^{-2}$)} &
    \colhead{} &
    \colhead{(min)} & \colhead{}
    }
  \startdata
  1 & & 15 & 11 & 1.1 & & 25 & 0.40\\
  2 & & 11 & 22 & 4.0 & & $\arcsec$ & $\arcsec$\\
  3 & & 10 & 8.6 & 1.9 & & 12 & 0.31\\
  4 & & 9.0 & 7.2 & 2.0 & & 6.8 & 0.27\\
  5 & & 13 & 15 & 2.0 & & 6.8 & 0.73\\
  6 & & 17 & 34 & 2.6 & & 12 & 0.74\\
  7 & & 23 & 27 & 1.1 & & 21 & 1.0\\
  \enddata
  \tablenotetext{a}{The lifetime and length
    of the dark surge
    are measured
    from the AIA 335 {\AA} slit-time diagram
    (Figure \ref{fig:slit}(b)),
    while the acceleration is calculated
    from the lifetime and length
    assuming a constant acceleration
    (see text for details).}
  \tablenotetext{b}{The duration
    and maximum normalized intensity
    of brightening
    measured from the AIA 1600 {\AA} lightcurve
    (Figure \ref{fig:slit}(d)).
    Events \#1 and 2 share
    common physical values
    since it is difficult
    to separate the two events.}
\end{deluxetable}




\end{document}